\documentclass[12pt]{iopart}

\usepackage{iopams} 
\usepackage{graphicx}
\usepackage[colorlinks]{hyperref}
\usepackage{cite}

\begin{document}

\title[Imaginary spin-orbital coupling in parity-time symmetric systems ...]{Imaginary spin-orbital coupling in parity-time symmetric systems with momentum-dependent gain and loss}

\author{Jieli Qin$^1$, Lu Zhou$^{2,4}$, Guangjiong Dong$^{3,4}$}

\address{$^1$ School of Physics and Materials Science, Guangzhou University, 230 Wai Huan Xi
    Road, Guangzhou Higher Education Mega Center, Guangzhou 510006, China}
\address{$^2$ Department of Physics, School of Physics and Electronic Science, East China
    Normal University, Shanghai 200241, China}
\address{$^3$ State Key Laboratory of Precision Spectroscopy, East China Normal University,
    Shanghai 200241, China}
\address{$^4$ Collaborative Innovation Center of Extreme Optics, Shanxi University, Taiyuan,
    Shanxi 030006, China}

\eads{\mailto{104531@gzhu.edu.cn}, \mailto{lzhou@phy.ecnu.edu.cn}, \mailto{gjdong@phy.ecnu.edu.cn}}
\vspace{10pt}
\begin{indented}
\item[] December 2021
\end{indented}

\begin{abstract}
Spin-orbital coupling (SOC) and parity-time ($\mathcal{PT}$) symmetry both
have attracted paramount research interest in condensed matter physics, cold
atom physics, optics and acoustics to develop spintronics, quantum computation,
precise sensors and novel functionalities. Natural SOC is an intrinsic
relativistic effect. However, there is an increasing interest in synthesized
SOC nowadays. Here, we show that in a $\mathcal{PT}$-symmetric
spin-1/2 system, the momentum-dependent balanced gain and loss can synthesize
a new type of SOC, which we call imaginary SOC. The imaginary SOC can
substantially change the energy spectrum of the system. Firstly, we show
that it can generate a pure real energy spectrum with a double-valleys
structure. Therefore, it has the ability to generate supersolid stripe states.
Especially, the imaginary SOC stripe state can have a high contrast of one.
Moreover, the imaginary SOC can also generate a spectrum with tunable complex
energy band, in which the waves are either amplifying or decaying. Thus,
the imaginary SOC would also find applications in the 
engineering of $\mathcal{PT}$-symmetry-based coherent wave
amplifiers/absorbers. Potential experimental realizations of imaginary SOC 
are proposed in cold atomic gases and systems of coupled waveguides 
constituted of nonlocal gain and loss.
\end{abstract}

%

\submitto{\NJP}

%
%

\section{Introduction} 
Spin-orbit coupling (SOC) arises from the
relativistic-induced coupling between a particle's spin degree of freedom and
its momentum. It plays a significant role in various physical systems. For
relativistic elementary particles, SOC leads to their Zitterbewegung
oscillation \cite{Schrodinger1930}. For atoms, SOC gives rise to the fine
structure spectra \cite{Bernath2005Spectra}. And in condensed matter physics,
the investigation of SOC has led to fruitful achievements (such as the
spin-Hall effect \cite{Sinova2015Spin}, topological insulator
\cite{Hasan2010Topological}, just name a few) with potential applications in
spintronics \cite{Zutic2004Spintronics} and quantum computations
\cite{Kloeffel2013Prospects}. Currently, SOC researches are drastically
expanding to the fields of cold atom physics
\cite{Zhai2012Spin,Galitski2013Spin,Zhai2015Degenerate,Zhang2016Properties,
    Zhang2019Recent,Lin2011Spin,Li2012Quantum,Hu2012Spin,Achilleos2013Matter,Li2017A,
    Bersano2019Experimental,Luo2019Tunable,Putra2020Spatial,Zhao2020Magnetic,
    Hou2018Momentum,Zheng2018Josephson}, optics
\cite{Bliokh2015Spin,Menyuk1989Pulse,Malomed1991Polarization,
    Chiang1997Experimental,Kartashov2015Dark,Kartashov2015Stabilization,
    Sakaguchi2016One} and acoustics
\cite{Ju2018Acoustic,Bliokh2019Spin,Deng2020Acoustic,Gao2020Acoustic,Wang2021Spin}.
Cold atom systems have
no natural SOC, whereas using the Raman coupling scheme \cite{Lin2011Spin}
(and also some others, see the review article \cite{Zhang2019Recent})
artificial SOC has been experimentally realized, furthermore supersolid stripe
states \cite{Li2012Quantum,Hu2012Spin,Achilleos2013Matter,Li2017A,
    Bersano2019Experimental,Luo2019Tunable,Putra2020Spatial,Zhao2020Magnetic} and
momentum-space Josephson oscillations
\cite{Hou2018Momentum,Zheng2018Josephson} can be generated. In photonics, the
subwavelength scales and additional degrees of freedom of structured optical
field are explored, and in such fields, spin and orbital properties are
strongly coupled with each other \cite{Bliokh2015Spin}. In a birefringent
optical fiber, traveling of the light pulses with mutually orthogonal linear
polarizations is described by a set of nonlinear Schr\"{o}dinger equations
with a SOC Hamiltonian \cite{Menyuk1989Pulse,Malomed1991Polarization}. In
dual-core waveguides, SOC can be synthesized by dispersively coupling the
light field in different cores \cite{Chiang1997Experimental,
    Kartashov2015Dark,Kartashov2015Stabilization}, and stripe solitons can be
produced \cite{Sakaguchi2016One}. In acoustical systems, SOC has also been
successfully synthesized \cite{Ju2018Acoustic,Bliokh2019Spin,Deng2020Acoustic,
Gao2020Acoustic,Wang2021Spin}.

Non-Hermitian parity-time ($\mathcal{PT}$) symmetry physics is another active
researching field these days
\cite{Bender2007Making,Christodoulides2018Parity,ElGanainy2018Non,Ozdemir2019Parity}%
. The seminal work by Bender and Boettcher in 1998 states that a non-Hermitian
Hamiltonian with $\mathcal{PT}$-symmetry can support complete real value
eigenenergies \cite{Bender1998Real}, and this has boosted the investigation of
non-Hermitian quantum mechanics
\cite{Bender2002Complex,Brody2016Consistency,Zhang2019Time,
    Lee2014Local,Quijandria2018PT,Wu2019Obervation}. Considering the analogy of
classical wave equation to the quantum mechanics Schr\"{o}dinger equation,
$\mathcal{PT}$-symmetric Hamiltonian can be used to model the balanced gain
and loss, and has been realized in systems of optics
\cite{Guo2009Oberservation,Ruter2010Oberservation}, acoustics
\cite{Zhu2014PT,Shi2016Accessing,Fleury2016Parity,Yang2019Experimental}, and
many others \cite{Schindler2011Experimental,Bittner2012PT,Hang2013PT,
Bender2013Oberservation,Peng2016Anti,Zhang2016Oberservation,Fang2021Universal}. 
Fruitful
applications such as invisible acoustic sensors \cite{Fleury2015An},
microcavity sensors \cite{Wiersig2014Enhancing,Yu2020Experimental},
unidirectional transportation \cite{Feng2013Experimental,Jin2017One,Gear2017Unidirctional},
light-light switch \cite{Zhao2016Metawaveguide}, and laser amplifier/absorber
\cite{Miri2012Large,Feng2014Single,Longhi2014PT,Longhi2010PT,Chong2011PT,Wan2011Time}
were born out.

Recently, researches on momentum-dependent gain and loss are emerging. In cold
atomic gas systems, the gain can be realized by injecting atoms into the
condensate using an atom laser \cite{Robins2008A}, while the loss can be
realized by exciting atoms firstly to an excited state with a laser beam and
then ejecting them out from the condensate via photon recoil
\cite{Li2019Oberservation}. Due to the momentum-distribution of the atom laser
\cite{Kohl2005Observing,Smith2019Engineering}, and the Doppler effect in
atom-light interaction \cite{Wuster2017Non-Hermitian}, gain and loss realized
in these ways have a distinct momentum-dependence. In the optical medium,
spatially nonlocal gain and loss after a Fourier transformation are
wavevector-dependent (or equivalently ``momentum''-dependent)
\cite{Raza2015Nonlocal,Xu2018Quantum}, and this feature has been proposed to
explore the topological physics in photonic systems \cite{Cerjan2018Effects}.

Here, we show that in $\mathcal{PT}$-symmetric spin-1/2 systems, 
momentum-dependent gain and loss can synthesize an imaginary 
interaction between the spin degree of freedom and the orbital motion, thus
an imaginary SOC (for comparison, the conventional SOC will be termed as ``real
SOC'' in the following). Next, we present the imaginary SOC
Hamiltonian, and study its properties at first, leaving the discussion on
potential experimental realizations in cold atom physics and optical systems
to the end of the paper.

\section{Model} 
We study a spin-1/2 Hamiltonian in
momentum representation given by%
\begin{equation}
    H=\frac{\left\vert\vec{p}\right\vert ^{2}}{2m}\sigma_{0}+i\hbar\gamma\Theta\left(  \vec{p}\right)  \sigma_{z}+\hbar\Omega\sigma_{x},
    \label{eq:Hamiltonian0}%
\end{equation}
where $\sigma_{0}$, $\sigma_{x,y,z}$ are the conventional $2\times2$ unitary 
and Pauli matrices, $m$ is the particle mass, $\vec{p}$ is the
momentum operator, $\hbar$ is the Planck constant, $\gamma$ represents the
rate of the balanced gain and loss, $\Theta(\vec{p})$ is a positive
dimensionless function with its maximum normalized to 1 [i.e., $0\leq
\Theta(\vec{p})\leq1$], and it is used to describe the profile of 
momentum-dependence of
the balanced gain and loss, at last $\Omega$ is the Rabi coupling strength
between the two spin components.

If $\Theta\left(  \vec{p}\right)  $ is imaginary, the second term in
Hamiltonian (\ref{eq:Hamiltonian0}) is real, and represents an interaction
between the spin and a momentum-dependent magnetic field along the
$z$-direction. Thus, it is a conventional real SOC term. For example, when
$\Theta\left(  \vec{p}\right)  =ip_{x}$, the Hamiltonian
(\ref{eq:Hamiltonian0}) has the same form as that in the SOC cold atom system
realized with the Raman coupling scheme \cite{Lin2011Spin}. However, in the
present paper, $\Theta\left(\vec{p}\right)$ is a profile function describing
the momentum-dependence of the gain and loss, that is to say it is a real
function, hence the second term of the Hamiltonian is imaginary, 
thus termed as ``imaginary SOC''. This
imaginary SOC Hamiltonian is $\mathcal{PT}$-symmetric, with the parity
operator $\mathcal{P}$ exchanging the two spin components, and the operator
$\mathcal{T}$ performing the complex conjugation \cite{Bender2007Making}.
We also note that there already exist some researches combining SOC and $\mathcal{PT}$
symmetry\cite{Kartashov2014CPT,Guo2021Theoretical,Zhou2021Engineering,
Sun2021Impurity,Watanabe2021Photocurrent},
however in these works it is a conventional Hermitian SOC term and a 
$\mathcal{PT}$-symmetric term been stiffly glued. In contrast, here
the SOC term itself is non-Hermitian.

\begin{figure}
    \begin{center}
        \includegraphics{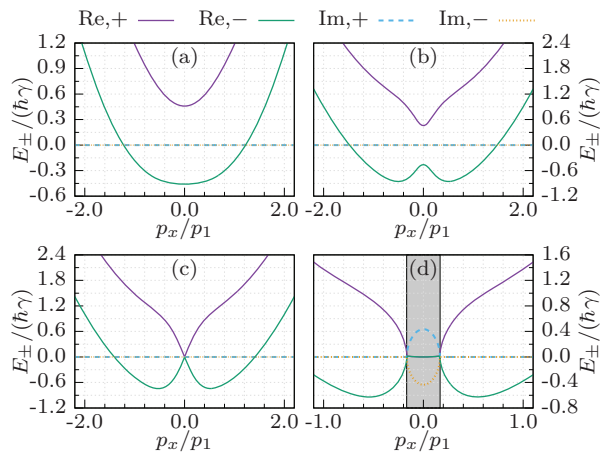}%
        \caption{Energy spectra $E_{\pm}\left(  p_{x}\right)  $ for imaginary SOC
            with momentum-dependence profile $\Theta\left(  p_{x}\right)  $ given by 
            Eq. (\ref{eq:Lorentzian}). The real (Re) parts of $E_{+}$ and $E_{-}$
            are plotted as the solid violet and green lines, while their imaginary (Im)
            parts are plotted as the cyan dashed and brown dotted lines. Panels (a,b):
            Pure real spectra for $\Omega=1.1\gamma>\gamma$. Under these parameters, 
            the critical gain/loss momentum-space width separates the single- 
            and double-valleys spectrum is $\sigma_{c}=2.09p_{1}$ with $p_{1}=\sqrt
            {m\hbar\gamma}$. Panel (a): Single-valley energy spectrum for $\sigma
            =2.5p_{1}>\sigma_{c}$. Panel (b): Double-valleys energy spectrum for
            $\sigma=0.5p_{1}<\sigma_{c}$. Panel (c): Pure real gapless energy spectrum for
            $\Omega=\gamma$ and $\sigma=0.5p_{1}$. Panel (d): Complex energy spectrum for
            $\Omega=0.9\gamma<\gamma$ and $\sigma=0.5p_{1}$. In the gray color
            filled band, the spectrum has complex eigenenergy.}%
        \label{fig:spectrum}%
    \end{center}
\end{figure}

To analytically understand the properties of imaginary SOC, we solve the
eigenvalue problem of Hamiltonian (\ref{eq:Hamiltonian0}). The eigenenergies and
corresponding eigenvectors are
\begin{equation}
    E_{\pm}=\frac{\left\vert \vec{p}\right\vert ^{2}}{2m}\pm\hbar\sqrt{\Omega
        ^{2}-\gamma^{2}\Theta^{2}\left(  \vec{p}\right)  }, \label{eq:Spectrum}%
\end{equation}
and%
\begin{equation}
    \psi_{\pm}=\frac{1}{A}
    \left[
    \begin{array}{c}
        \left(  i\gamma\Theta\left(\vec{p}\right)\pm\sqrt{\Omega
            ^{2}-\gamma^{2}\Theta^{2}\left(  \vec{p}\right)  } \right)  /\Omega\\
        1
    \end{array}
    \right],
     \label{eq:Eigenstates}%
\end{equation}
with $A_{\pm}$ being normalization constants. When $\Omega\geq\gamma$, the
eigenenergies $E_{\pm}$ always take real values, and the eigenvectors can be
simplified to $\psi_{\pm}=\left[\pm \exp\left(\pm i\theta\right),
1\right] ^{T}/\sqrt{2}$
with $\theta=\arcsin\left[\gamma\Theta\left(  \vec{p}\right)  /\Omega\right]$,
and ``T'' denoting the transpose operation. It is seen that these eigenvectors
are $\mathcal{PT}$-symmetric with equal spin amplitudes $\left\vert \psi_{\pm,\uparrow
}\right\vert =\left\vert \psi_{\pm,\downarrow}\right\vert $. Considering the
dynamical factor $\exp\left[  -iE_{\pm}t/\hslash\right]  $, the states will
evolve with their norms conserved due to the exact balance between loss and gain.
When $\Omega<\gamma$, the states with momentum fulfilling $\gamma^{2}\Theta
^{2}\left(  \vec{p}\right)  \leq\Omega^{2}$ are also $\mathcal{PT}$-symmetric
with pure real eigenenergies. However, the states with momentum fulfilling
$\gamma^{2}\Theta^{2}\left(  \vec{p}\right)  > \Omega^{2}$ have complex
eigenenergies $E_{\pm}=E_{R}\pm iE_{I}$ with $E_{R}=\left|\vec{p}\right|^{2}/\left(
2m\right)  $ and $E_{I}=\hbar\sqrt{\gamma^{2}\Theta^{2}\left(  \vec{p}\right)
    -\Omega^{2}}$. Now, the dynamical factor becomes $\exp\left[  -iE_{\pm
}t/\hslash\right]  =\exp\left[  -iE_{R}t/\hbar\right]  \cdot\exp\left[  \pm
E_{I}t/\hbar\right]  $, for $\exp\left[  -E_{I}t/\hbar\right]  $ the states
will automatically decay, while for $\exp\left[  +E_{I}t/\hbar\right]  $ the
states will automatically be amplified during the evolution. The decay and
amplify can be understood by examining the corresponding eigenvectors. In this
case, the eigenvectors are simplified to $\psi_{\pm}=\left[i\exp\left(\pm\vartheta
\right),1\right]^{T}/\sqrt{A_\pm}$ with $\vartheta=\mathrm{arccosh}\left[
\gamma\Theta\left(\vec{p}\right)/\Omega\right]$ and $A_\pm = \sqrt{1+
    \exp\left(\pm 2 \vartheta\right)}$, which have unequal spin
amplitudes $\left\vert \psi_{\pm,\uparrow}\right\vert \neq\left\vert \psi
_{\pm,\downarrow}\right\vert $. Thus, the $\mathcal{PT}$-symmetry is broken,
the gain and loss no longer can cancel each other.

Next, we examine the effect of imaginary SOC on the shape of energy spectrum
curve. For facilitation, we particularly consider the one-dimensional case
($\vec{p}=p_{x}\hat{x}$), and assume that the momentum-dependence profile 
function $\Theta\left(  p_{x}\right)  $ takes the Lorentzian shape
\begin{equation}
    \Theta\left(  p_{x}\right)  =\frac{\sigma^{2}}{p_{x}^{2}+\sigma^{2}},
    \label{eq:Lorentzian}%
\end{equation}
where $\sigma$ is the momentum-space width of the balanced gain and loss.

Firstly, we found that the imaginary SOC can generate spectra with both
single-valley and double-valleys structures when $\Omega\geq\gamma$, as shown
in panels (a,b) of Fig. \ref{fig:spectrum}. The minima and maxima of the
spectrum can be found by letting the derivatives of 
$E_{\pm}\left(p_x\right)$ equal to zero,
$dE_{\pm}/dp_{x}=0$. The upper branch $E_{+}$ of the spectrum always has only
one minimum at $p_{x}=0$. However, the lower branch $E_{-}$ can have either
one or two minima, depending on the value of $\sigma$. When $\sigma$ is
larger than the critical value $\sigma_{c}=\sqrt{2m\hbar} \gamma/\left(
\Omega^{2}-\gamma^{2}\right)  ^{1/4}$, the lower branch also has only one
minimum at $p_{x}=0$. While for $\sigma<\sigma_{c}$, the point $p_{x}=0$
becomes a maximum, and the lower branch spectrum exhibits a double-valleys
structure with two minima located at $p_{x}=\pm p_{0}$, which are the
solutions of equation $dE_{-}/dp_{x}=0$.

Usually, there is an energy gap between the upper and lower spectrum branches,
$\Delta E=2\hbar\sqrt{\Omega^{2}-\gamma^{2}}$. However, due to the energy band
attraction effect in non-Hermitian systems \cite{Harder2018Level,Wu2020Energy}%
, as the strength of the balanced gain and loss increases, the gap becomes
narrow, and when $\gamma=\Omega$, it disappears, see panel (c) of Fig.
\ref{fig:spectrum}. Interesting, even the spectrum becomes gapless, the two
spin components are still coupled with each other [see eigenstates
(\ref{eq:Eigenstates})], in sharp contrast with the real SOC case where the
gapless system is trivial due to the decoupling of the two spin components
\cite{Lin2011Spin,Achilleos2013Matter,Zhou2015Goos,Qin2020Bound,
    Qin2020Unidirectional,Chen2020Spin}.

We also found that the imaginary SOC can generate spectrum with a complex
energy band when $\Omega<\gamma$, as shown by the gray color filled rectangle
in panel (d) of Fig. \ref{fig:spectrum}. 
In this case, there exist a critical momentum
$p_{c}=\sigma\sqrt{\gamma/\Omega-1}$ [the solution of $\Omega^{2}-\gamma
^{2}\Theta^{2}\left(  p_{x}\right)  =0$], the state with momentum $\left\vert
p_{x}\right\vert \geq p_{c}$ still has pure real eigenenergy. However, the
eigenenergy corresponding to state with momentum $\left\vert
p_{x}\right\vert <p_{c}$ becomes complex, thus forming the complex energy band
in the figure. And according to formula $p_{c}=\sigma\sqrt{\gamma/\Omega-1}$,
this complex energy band can be conveniently tuned by both the gain/loss or
Rabi coupling strength.

The above discussion indicates that imaginary SOC would find 
applications in the fields of energy spectrum engineering and 
coherent wave amplifying/absorbing. Similar to real
SOC, imaginary SOC also can generate spectra with single-valley or double
valleys structures. Additionally, imaginary SOC also generates non-trivial
gapless spectrum and spectrum with complex energy band, which are absent in
the case of real SOC. It has been proposed and experimentally demonstrated that
the complex energy modes of a $\mathcal{PT}$-symmetric optical
system can be applied to realize laser amplifiers and absorbers
\cite{Miri2012Large,Feng2014Single,Longhi2014PT,Longhi2010PT,Chong2011PT,Wan2011Time}. 
Here we show that imaginary SOC can control the complex energy spectrum,
thus provides it the ability to manage $\mathcal{PT}$-symmetry-based laser 
amplifiers/absorbers.%

\begin{figure}
    \begin{center}
        \includegraphics{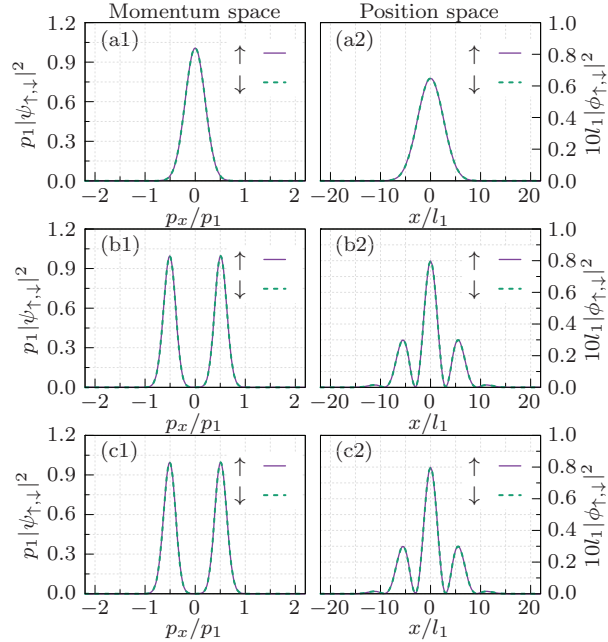}
        \caption{$\mathcal{PT}$-symmetric normal and supersolid stripe eigenstates
            generated by imaginary SOC in a harmonic trap. The left panels (a1-c1) show
            the probability density in momentum space, while the right panels (a2-c2) show
            the probability density in position space. The spin-$\uparrow$ and
            spin-$\downarrow$ componets are plotted using violet solid and green dashed
            line respectively (since the states are $\mathcal{PT}$-symmetric with
            $\left|\psi_\uparrow\right|=\left|\psi_\uparrow\right|$, the two lines 
            overlap with each other). The harmonic trap frequency is 
            $\omega=0.05\gamma$ for all the
            panels. Panles (a1, a2): Normal state under parameters $\Omega=1.1\gamma$,
            $\sigma=2.5p_{1}$ ($p_{1}=\sqrt{m\hbar\gamma}$). Panels (b1, b2): Supersolid
            Stripe state under parameters $\Omega=1.1\gamma$, $\sigma=0.5p_{1}$. Panels
            (c1, c2): Supersolid Stripe state under parameters $\Omega=\gamma$,
            $\sigma=0.5p_{1}$. The pure real eigenenergies of these states are
            $E=-0.44\hbar\gamma$ (a), $-0.81\hbar\gamma$ (b), $-0.70\hbar\gamma$ (c). The
            corresponding free energy spectra are shown in panels (a, b, c) of figure
            \ref{fig:spectrum}.}%
        \label{fig:NormalAndStripeState}%
    \end{center}
\end{figure}

\section{Imaginary SOC in Harmonic Trap}
For the single-valley
spectrum, it is easy to imagine that the ground state of a imaginary SOC system
will fall into the
bottom of this valley. However, for the double-valleys spectrum, the ground
state may fall in either the $p_{0}$ valley or the opposite $-p_{0}$ valley.
Thus, when an external trapping potential is included, 
the superposition of these two
momentum states can be induced to produce a supersolid stripe state. In this
vein of thought, now we consider a case that the 
imaginary SOC particles are trapped in a harmonic
potential $V(x)=\frac{1}{2}m\omega^{2}x^{2}$, or equivalently $V(p_{x}%
)=-\frac{1}{2}m\hbar^{2}\omega^{2}\partial_{p_{x}}^{2}$ in the momentum
representation. We first numerically diagonalize the full Hamiltonian 
in the momentum representation to obtain the eigenwavefunctions $\psi
_{n}(p_{x}) = \left[\psi_{\uparrow,n}(p_{x}),\psi_{\downarrow,n}%
(p_{x})\right]  ^{T}$ with $n=1,2,\cdots$, and
further calculate the wavefunction in position representation by Fourier
transform, $\phi_{n}(x)=\int\psi_{n}(p_{x})e^{ip_{x}x/\hbar}dx/\sqrt{2\pi}$.

In Fig. \ref{fig:NormalAndStripeState}, we show the probability densities of
some different eigenstates in both momentum and position space for a weak harmonic trap
with frequency $\omega=0.05\gamma$. Other parameters are chosen corresponding
to the free spectra previously shown in panels (a,b,c) of Fig. 
\ref{fig:spectrum}. Panels (a1,a2) correspond to the single-valley
spectrum. We see that in momentum space the probability density is centered at
$p_{x}=0$, which is just the center of the spectrum valley; and in position
space, the density is centered at the bottom of the harmonic trap (since it is
similar to the ground state of a normal harmonic oscillator, we call it
``normal'' state). When the free spectrum has a double-valleys structure, we
found the probability density of the bound state exhibits two peaks
centered around the bottom of the two valleys; and in position space, the
expected supersolid stripe structure is observed (``stripe'' state), see
panels (b1,b2; c1,c2). Especially, the free spectrum corresponding to panels
(c1,c2) is gapless, but the stripe state is also observed. This further
demonstrates the non-trivialness of the gapless free spectrum which has been
discussed in the previous section. More interesting, we see that the density minimums
of the imaginary SOC generated stripe state can drop to zero, i.e.,
the stripe has a high contrast of one. By comparison, the supersolid stripe
realized with the real SOC usually has a poor contrast, leading to difficulty in
experimental observation
\cite{Martone2014Approach,Martone2015Visibility,Chen2018Quantum}.

\begin{figure}
    \begin{center}
        \includegraphics{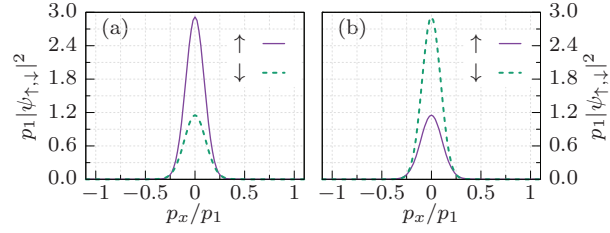}%
        \caption{$\mathcal{PT}$-symmetry broken eigenstates generated by imaginary SOC
            in a harmonic trap. The probability in momentum space for spin-$\uparrow$
            and spin-$\downarrow$ components are plotted as the violet solid and green
            dashed lines respectively. Panel (a): Decaying eigenstate with complex
            eigenenergy $E=\left(  0.08-0.36i\right)  \hbar\gamma$. Panel (b): Amplifying
            eigenstate with complex energy $E=\left(  0.08+0.36i\right)  \hbar\gamma$. The
            parameters used are $\omega=0.05\gamma$, $\Omega=0.9\gamma$ and $\sigma
            =0.5p_{1}$ ($p_{1}=\sqrt{m\hbar\gamma})$. The corresponding free energy
            spectrum has
            a complex energy band around $p_{x}=0$, as shown in panel (d) of Fig.
            \ref{fig:spectrum}.}%
        \label{fig:PTBrodenState}%
    \end{center}
\end{figure}

Besides, corresponding to the complex energy band free spectrum [panel (d) of
Fig. \ref{fig:spectrum}], imaginary SOC can also generate $\mathcal{PT}%
$-symmetry broken bounded states in the harmonic trap. The result is shown in
Fig. \ref{fig:PTBrodenState}. The $\mathcal{PT}$-symmetry broken states appear
in pairs with opposite spin polarization and opposite imaginary parts of
eigenenergies, thus one of them is a decaying state, while the other is an
amplifying one.

\section{Experimental Realizations}
Firstly, we
propose that imaginary SOC can be realized in cold atom systems. We consider a
system of one-dimensional spin-1/2 cold atomic gas with the two spin
components coupled by a laser field with Rabi frequency $\Omega$. The
momentum-dependent gain in the spin-$\uparrow$ component can be realized by
injecting spin-$\uparrow$ atoms into the system using an atom laser with
appropriate momentum distribution
\cite{Robins2008A,Kohl2005Observing,Smith2019Engineering}. Especially, ref.
\cite{Smith2019Engineering} proposed a procedure to produce cold-atom beams
with the Lorentzian profile of momentum distribution. And according to ref.
\cite{Wuster2017Non-Hermitian}, exploiting the Doppler shift technique, atoms
in the spin-$\downarrow$ state can be velocity-selectively excited to narrow
Rydberg or metastable states with a laser, and the following photon recoil
could produce a momentum-dependent atomic loss. Denoting the
momentum-dependent gain and loss by $\pm i\hbar\gamma\Theta\left(
\vec{p}\right)  $, the system can be readily described by the Hamiltonian
(\ref{eq:Hamiltonian0}).

Another possible system to realize the imaginary SOC is the coupled planar
waveguides with nonlocal gain and loss \cite{Raza2015Nonlocal,Xu2018Quantum}.
We consider a system of two coupled planar ($x$-$z$ plane) waveguides, where
light propagates along the $z$-axis, and $x$ is the transverse direction. In
the paraxial approximation, the light fields in the two waveguides
$E_{\uparrow,\downarrow}$ follow equations
\cite{Agrawal2001Nonlinear,Makris2008Beam,Guo2009Oberservation,Ruter2010Oberservation}%
\begin{eqnarray}
    i\frac{\partial E_{\uparrow,\downarrow}\left(  z,x\right)  }{\partial z}=  &
    -\frac{\lambda_{0}}{4\pi n_{0}}\frac{\partial^{2}E_{\uparrow,\downarrow
        }\left(  z,x\right)  }{\partial x^{2}}+\Omega E_{\downarrow,\uparrow}\left(
    z,x\right) \nonumber\\
    &  \mp\frac{2\pi}{\lambda_{0}}\int i\gamma\Xi\left(  x-x^{\prime}\right)
    E_{\uparrow,\downarrow}\left(  z,x^{\prime}\right)  dx^{\prime},
    \label{eq:Eupdownx}%
\end{eqnarray}
where $\lambda_{0}$ is the vacuum wavelength, $n_{0}$ is the background
refractive index of the waveguides, $\Omega$ is the coupling between the two
waveguides, and $\pm i\gamma\Xi\left(  x-x^{\prime}\right)  $ describes the
nonlocal gain and loss in the two waveguides. Performing Fourier transform on
both sides of Eqs. (\ref{eq:Eupdownx}), we get
\begin{equation}
    i\frac{\partial}{\partial z}\left[
    \begin{array}{c}
        E_{\uparrow}\left(  z,k_{x}\right) \\
        E_{\downarrow}\left(  z,k_{x}\right)
    \end{array}
    \right]  =\mathcal{H}\left[
    \begin{array}{c}
        E_{\uparrow}\left(  z,k_{x}\right) \\
        E_{\downarrow}\left(  z,k_{x}\right)
    \end{array}
    \right]  , \label{eq:EupEdown}%
\end{equation}
with
\begin{equation}
    \mathcal{H=}-\frac{\lambda_{0}k_{x}^{2}}{4\pi n_{0}}\sigma_{0}+\frac{2\pi
    }{\lambda_{0}}i\gamma\Theta\left(  k_{x}\right)  \sigma_{z}+\Omega\sigma_{x},
    \label{eq:EHamiltonian}%
\end{equation}
where $\Theta\left(\cdot\right)$ is the Fourier transform of
function $\Xi\left(\cdot\right)$. Now, it is obvious that Eq.
(\ref{eq:EupEdown}) is analogous to an Schr\"{o}dinger equation in momentum
representation, where the ``Hamiltonian'' $\mathcal{H}$ has the same form
as that of Eq. (\ref{eq:Hamiltonian0}), with wavevector $k_x$ playing the role of
momentum. Therefore, we propose that
the coupled optical waveguides system with nonlocal gain and loss can be used
to emulate the imaginary SOC.

\section{Summary}
In this work, we have proposed a new type of SOC---imaginary SOC, 
which is induced by momentum-dependent gain and loss, and possesses
the property of $\mathcal{PT}$-symmetry.
Imaginary SOC has the ability of energy spectrum engineering. As examples, we
firstly show that it can generate pure real energy spectra with
double-valleys structures. The interference between waves corresponding to
different valleys can produce supersolid stripe states. Compared to the real
SOC supersolid stripe state which usually has low contrast, the
supersolid stripe state realized by imaginary SOC can have a high contrast 
of one. What is more, we also shown that the imaginary SOC
can generate spectra with tunable band
of complex energy, in which the waves are either decaying or amplifying
during the time evolution. Thus, the imaginary SOC can also have potential
applications in engineering the $\mathcal{PT}$-symmetry-based coherent wave
amplifiers/absorbers. For experimental realization, we propose
that the imaginary SOC can be implemented in spin-1/2 cold atomic gases
and also in systems of coupled waveguides.

\section*{Acknowledgments}
J. L. Qin acknowledges the support from National Natural Science Foundation of
China (11904063). G. J. Dong acknowledges the financial support from Shanghai
Municipal Education Commission (2019-01-07-00-05-E00079), National Natural
Science Foundation of China (11574085, 11834003, 91536218), National Key
Research and Development Program of China (2017YFA0304201) and Overseas
Expertise Introduction Project for Discipline Innovation (B12024). L. Zhou
acknowledges the support from National Natural Science Foundation of China
(12074120) and Science and Technology Commission of Shanghai Municipality
(Grant No. 20ZR1418500).

\end{document}